\documentclass[9pt,a4paper]{extarticle}
\usepackage{multicol}
\usepackage[T1]{fontenc}
\usepackage[utf8]{inputenc}
\usepackage{lmodern,microtype,needspace}
\usepackage[margin=0.61in]{geometry}
\usepackage{amsmath,amssymb,amsthm,mathtools}
\usepackage{booktabs,tabularx,array,enumitem}
\usepackage{xcolor}
\definecolor{ink}{HTML}{193247}
\definecolor{teal}{HTML}{117C83}
\usepackage[colorlinks=true,linkcolor=ink,citecolor=teal,urlcolor=teal]{hyperref}
\usepackage{fancyhdr}
\setlist{nosep,leftmargin=*}
\newtheorem{proposition}{Proposition}

\newcolumntype{L}[1]{>{\raggedright\arraybackslash}p{#1}}
\newcommand{\E}{\mathbb E}
\newcommand{\Var}{\operatorname{Var}}

\newcommand{\sgn}{\operatorname{sgn}}
\newcommand{\dd}{\mathrm d}

\newcommand{\CE}{\operatorname{CE}}

\newcommand{\sectionstanza}[1]{%
  \par\vspace{-0.55em}%
  \begin{flushright}%
    \begin{minipage}{0.54\linewidth}%
      \raggedright\fontsize{7.80}{8.6}\selectfont\itshape
      #1
    \end{minipage}%
  \end{flushright}%
  \vspace{-0.45em}%
}
\newcommand{\stanzaindent}{\hspace*{1.2em}}
\newcommand\blfootnote[1]{%
  \begingroup
  \renewcommand\thefootnote{}\footnote{#1}%
  \addtocounter{footnote}{-1}%
  \endgroup
}

\title{\textbf{Generalised Market Impact for Equities and Derivatives}\\[1pt]}
\author{Bernhard K Meister}
\date{\today}

\begin{document}
\maketitle
\thispagestyle{empty}

\noindent\hspace*{\fill}\textit{The buyer needs a hundred eyes, the seller not one.}\\[2pt]
\hspace*{\fill}--- George Herbert, \textit{Jacula Prudentum} (1651)

\vspace{0.0005em}

\begin{abstract}
\noindent
Market impact is often discussed from the perspective of the liquidity taker, whose orders move the price. We take the perspective of the liquidity provider, who must decide at what price to accept the resulting inventory. Under a mean-variance criterion, we derive how the provider's quote adjustment scales with position size. The resulting exponent depends on three inputs. Two govern the scaling of incremental portfolio variance with position size and time. The third relates the anticipated holding period to position size.
The model extends to options and shows how inventory risk can contribute to the asymmetry between overnight and intraday returns.

\end{abstract}

\section{Introduction}
 \sectionstanza{
``The time has come,'' the Walrus said,\\
\stanzaindent ``To talk of many things:\\
Of shoes---and ships---and sealing-wax---\\
\stanzaindent Of cabbages---and kings---\\
And why the sea is boiling hot---\\
\stanzaindent And whether pigs have wings.''
}

\label{sec:intro}
\blfootnote{Stanzas are from Lewis Carroll's \emph{The Walrus and the Carpenter} (1871).}

How much should a liquidity provider charge for an incremental position change?
The answer depends on inventory, capital and expected holding period.
Customers trade for many reasons---information, cash-flow needs or institutional schedules---while for a liquidity provider (LP) accommodating orders is the business.
A transaction records agreement on a price, not necessarily agreement on value. 
The familiar square-root relation
between price displacement and traded quantity is a useful empirical
benchmark \cite{Toth2011,Donier2015}, but need not describe a single provider's supply-price schedule.
A second-order log-utility calculation yields a more general three-exponent relation.




Section~\ref{sec:lp} sets out the individual portfolio problem;
Section~\ref{sec:exponent} derives the exponent and its admissible
range. Sections~\ref{sec:options} and~\ref{sec:knut} apply the result
to volatility-surface risk and to overnight-intraday price reversals. Section~\ref{sec:scope}
summarises the results, while
Appendix~\ref{app:growth} relates the quadratic criterion to expected log growth
and discusses its limits.

\section{The Liquidity Provider's Toil}

\par\smallskip
\begin{minipage}[t]{0.48\linewidth}
\vspace{0pt}\footnotesize\itshape
``O Oysters, come and walk with us!''\\
\stanzaindent The Walrus did beseech.\\
``A pleasant walk, a pleasant talk,\\
\stanzaindent Along the briny beach:\\
We cannot do with more than four,\\
\stanzaindent To give a hand to each.''
\end{minipage}\hfill
\begin{minipage}[t]{0.48\linewidth}
\vspace{0pt}\footnotesize\itshape
The eldest Oyster looked at him,\\
\stanzaindent But never a word he said:\\
The eldest Oyster winked his eye,\\
\stanzaindent And shook his heavy head---\\
Meaning to say he did not choose\\
\stanzaindent To leave the oyster-bed.
\end{minipage}\par\medskip

\label{sec:lp}

Consider one liquidity provider (LP) with wealth $W>0$ and an existing 
 portfolio. During an incremental inventory decision, $W$
is a state variable rather than a function of position size.

Let $p$ denote the current asset price and $K$ the LP's working estimate
of value, held fixed over the decision horizon. For a position bought
below this estimate, put $m=K-p>0$. The LP forecasts a price-change rate per unit of
\begin{equation}
 \mu=\kappa(K-p)=\kappa m,\qquad \kappa>0.
 \label{eq:drift}
\end{equation}
This is the restoring drift of the fractional  Ornstein–Uhlenbeck (OU) model $\mathrm dp_t=\kappa(K-p_t)\mathrm dt+\sigma\mathrm dB_t^H$, with ordinary OU dynamics when $H=1/2$. Here the drift serves as a local price-change forecast, not as a model of the price response to the LP's own trades. The reference level $K$ is an estimate rather than an objectively known fair value.


Suppose that after acquiring $q>0$ units and rebalancing its hedges, the LP's book has terminal   P\&L $Z_T(q)$ over the holding horizon $T$.  Under ordinary OU dynamics, holding $q$ units for
time $T$ gives an expected price gain of $qm(1-e^{-\kappa T})$.
The local drift forecast uses $q\kappa mT$ for the gain attributable
to the concession. Under ordinary OU dynamics this is a first-order
approximation, so the scaling result applies over inventory sizes for
which $\kappa T(q)\ll 1$. Expected hedging carry is included in the
reference forecast. The \emph{incremental} conditional risk is written as
\begin{equation}
 \mathcal V(q,T)=\Var_t[Z_T(q)]-\Var_t[Z_T(0)].
 \label{eq:varincrement}
\end{equation}
This measures the change in the variance of the LP's book, including the covariance of the trade and its hedges with existing positions. Trades are considered for which this change is positive.
For fixed hedges and covariance matrices, $\mathcal V$ is ordinarily
linear-plus-quadratic in $q$, so a fractional power is a
portfolio-level scaling assumption over a finite range, not a consequence of the quadratic criterion itself.

Over a finite range of inventory sizes and horizons, it is  assumed, 
\begin{equation}
\mathcal V(q,T)=V_*
       \left(\frac{q}{q_*}\right)^{2-\beta}
       \left(\frac{T}{T_*}\right)^{2H},
 \quad V_*,q_*,T_*>0,\quad 0\le\beta<1,\quad 0<H<1.
 \label{eq:riskscaling}
\end{equation}
Here \(q_*\) is a reference inventory size for the LP, \(T_*\) is a reference holding horizon, and \(V_*\) is the variance increment at \((q_*,T_*)\).
For the inventory decision considered here, the LP takes \(\beta\) as given. This exponent describes how incremental variance scales with position size across the \emph{whole rebalanced book}, reflecting netting, diversification and the LP’s capital constraints. Over time, the LP’s quotes may affect the order flow it attracts and thereby change the book’s composition and risk.
The quadratic risk charge may be written using a notional effective risk
capacity $\mathcal C(q)=W(q/q_*)^\beta$ in place of fixed capacity $W$;
$\mathcal C$ is not the LP's actual wealth, which remains fixed during
the decision\footnote{The split is a gauge choice: only the risk-charge exponent
$2-\beta$ is identified, since $\mathcal V\propto q^{x}$ and
$\mathcal C\propto q^{\,x-(2-\beta)}$ give the same certainty equivalent
for any $x$; the $x=2$ anchor identifies $\beta$ with the shortfall of
the variance exponent from the mechanical $q^2$ baseline.}. Fixed effective capacity gives $\beta=0$, and
$\mathcal C\propto\sqrt q$ gives $\beta=1/2$.  No value of $\beta$ is implied by the quadratic criterion. A hard, binding capital
limit instead truncates the supply schedule, and need not be represented
by a power law at all. 
The time exponent $H$  is that of the residual hedged P\&L\footnote{
For a fixed holding of \(q\) shares in one stock, considered in isolation, the P\&L is \(q\,\Delta p\) and inherits the time-scaling exponent of the stock’s price changes. In a portfolio, the corresponding exponent depends on how the variances and covariances of the residual exposures scale with time.
}.  

The LP's incremental `certainty equivalent' is defined over the holding horizon $T$ as the expected gain $q\kappa mT$ less a risk charge $\mathcal V(q,T)/(2W)$:
\begin{equation}
 \CE(q;m,T)=q\kappa mT-\frac{\mathcal V(q,T)}{2W}.
 \label{eq:CE}
\end{equation}
An LP typically adds a small position to an existing book. When netting
and hedging keep the book's P\&L small relative to wealth over the
holding horizon, both before and after the trade, \eqref{eq:CE} gives
the second-order approximation to the change in the LP's log-utility
certainty equivalent.
The term $q\kappa mT$ is the LP's forecast of the incremental gain.
For ordinary OU it is a short-horizon approximation; for fractional OU
with $H\ne\tfrac12$ we assume this forecast separately, since the
conditional mean can depend on the observed price history.
Appendix~\ref{app:growth} gives the expansion and its limitations.

\section{The Impact Exponent }
\sectionstanza{
Four other Oysters followed them,\\
\stanzaindent And yet another four;\\
And thick and fast they came at last,\\
\stanzaindent And more, and more, and more---\\
All hopping through the frothy waves,\\
\stanzaindent And scrambling to the shore.
}
\label{sec:exponent}

For a given $T$, maximising \eqref{eq:CE} with respect to $q$ yields 
\begin{equation}
 \kappa mT=\frac{(2-\beta)V_*}{2Wq_*}
   \left(\frac{q}{q_*}\right)^{1-\beta}
   \left(\frac{T}{T_*}\right)^{2H}.
 \label{eq:focfixed}
\end{equation}
Equivalently, the discount required by the LP  to hold $q$ units
is proportional to
\begin{equation}
 m(q;T)\ \propto\ q^{1-\beta}T^{2H-1}.
 \label{eq:inverse}
\end{equation}
The calculation is entirely local to the LP's own book: neither its
market share nor the total volume traded appears.

The LP quotes against the time it expects to carry inventory before offsetting flow arrives. This horizon depends on market conditions. 
The model takes three scaling exponents as inputs. In \eqref{eq:riskscaling}, $H$ governs the time dependence of incremental portfolio variance and $\beta$ its dependence on inventory size. The third exponent, $\gamma$, describes how the characteristic holding horizon under prevailing market conditions varies with inventory size,
\begin{equation}
 T(q)=T_*\left(\frac{q}{q_*}\right)^\gamma,
 \qquad\gamma\ge0.
 \label{eq:horizon}
\end{equation}
One may optimise at fixed $T$ and then compare decisions with
different horizons; if instead \eqref{eq:horizon} constrains the LP's
actual choice and is differentiated within its objective, the coefficient
changes but the power of $q$ does not.

\begin{proposition}[Individual LP scaling law]
\label{prop:law}
Suppose the LP maximises  \eqref{eq:CE}, with portfolio risk given by \eqref{eq:riskscaling} and holding horizon given by \eqref{eq:horizon}.
When the LP accounts for $T(q)$ in its choice of
$q$, the unrestricted objective has a unique interior maximum for each
$m>0$ precisely when $\delta>0$, with $\delta$ defined below. In that
case the inverse optimal-inventory schedule is
\begin{equation}
m(q)=A\left(\frac{q}{q_*}\right)^\delta,
 \qquad\delta=1-\beta+\gamma(2H-1),
 \label{eq:law}
\end{equation}
where, when the LP differentiates the entire horizon-dependent objective,
\begin{equation}
 A=\frac{V_*}{2\kappa Wq_*T_*}
       \frac{2-\beta+2H\gamma}{1+\gamma}>0.
 \label{eq:constant}
\end{equation}
The same exponent obtains when $T$ is held fixed during each individual
decision and the horizon schedule is applied afterward; only the factor
$(2-\beta+2H\gamma)/(1+\gamma)$ is then replaced by $2-\beta$.
\end{proposition}
\begin{proof}
With $x=q/q_*$ and \eqref{eq:horizon}, the first term of \eqref{eq:CE} is
proportional to $x^{1+\gamma}$ and its risk charge to
$x^{2-\beta+2H\gamma}$. Differentiation gives
\begin{equation*}
 \kappa m(1+\gamma)T
   =\frac{V_*}{2Wq_*}(2-\beta+2H\gamma)
     x^{1-\beta}\left(\frac{T}{T_*}\right)^{2H}.
\end{equation*}
Dividing by $T$ and inserting \eqref{eq:horizon} shows both
\eqref{eq:law} and \eqref{eq:constant}.
\end{proof}

For a fixed concession \(m>0\) and holding horizon \(T(q)\), the certainty equivalent becomes
\begin{equation}
 \CE(q)=a q^{1+\gamma}-b q^{2-\beta+2H\gamma},
 \qquad a,b>0,
 \label{eq:growthpowers}
\end{equation}
and a unique finite positive optimum exists precisely when the second
exponent exceeds the first, that is, when $\delta>0$. For $\delta<0$ the
interior stationary point is a minimum and the objective grows without
bound as $q$ increases. At $\delta=0$ the objective reduces to
$(a-b)q^{1+\gamma}$. It grows without bound if $a>b$, is maximised at
$q=0$ if $a<b$, and is constant if $a=b$. These cases describe the
unrestricted objective; the assumed power laws apply over a finite range.

A commonly observed impact regime is increasing and concave. Since
\eqref{eq:law} is increasing for $\delta>0$ and strictly concave for $0<\delta<1$,
its admissible parameter region is
\begin{equation}
 0<\delta<1\quad\Longleftrightarrow\quad
 \beta-1<\gamma(2H-1)<\beta.
 \label{eq:bounds}
\end{equation}
For $0\le\beta<1$ and $\gamma\ge0$, rough risk ($H<1/2$) requires
$\gamma<(1-\beta)/(1-2H)$; persistent risk ($H>1/2$) requires
$\gamma<\beta/(2H-1)$. At $H=1/2$ the horizon exponent cancels and
$\delta=1-\beta$. The boundary $\delta=1$ is a linear supply schedule,
not a forbidden market state.

Two checks are useful. With $(\beta,\gamma)=(1/2,1)$ one obtains
$\delta=2H-1/2$, the fractional special case discussed in
\cite{Meister2026}; with $H=1/2$ and $\beta=1/2$ one recovers the
square-root exponent. Special cases do not explain why the chosen
risk-scaling exponent must hold in a particular market.

\paragraph{From an LP quote to observed impact.}
Equation~\eqref{eq:law} is the LP's \emph{reservation-price schedule},
not a price-formation model. It describes observable marginal impact only
insofar as this provider's quote is relevant to the transaction and $K$
is locally unchanged. If $Q$ denotes aggregate market flow and this
provider accepts $q=\varphi(Q)Q$, a size-independent share $\varphi$
preserves the power exponent\footnote{The empirical square-root law expresses the metaorder size \(Q\) as a fraction of the asset's daily traded volume \(V_D\). If the LP takes a constant share \(\varphi\) of each metaorder, choosing the inventory scale \(q_*=\varphi V_D\) gives \(q/q_*=Q/V_D\). This choice makes the volume scales directly comparable. The impact coefficient still requires separate calibration.}; a varying share need not. Neither $Q$, the
market's average daily volume, nor $\varphi$ is determined by the
individual optimisation, and feedback from collective rebalancing to the
market price is a separate consideration \cite{Meister2016}.

\section{Application I: Options and Volatility-Surface Risk}
\sectionstanza{
``A loaf of bread,'' the Walrus said,\\
\stanzaindent ``Is what we chiefly need:\\
Pepper and vinegar besides\\
\stanzaindent Are very good indeed---\\
Now if you're ready, Oysters dear,\\
\stanzaindent We can begin to feed.
}
\label{sec:options}

An option LP typically manages a portfolio rather than an isolated option position: delta is hedged
in the underlying, and vega, skew, curvature, and term-structure
exposures are often offset across strikes and maturities. To first order,
\begin{equation}
 \dd C\simeq \frac{\partial C}{\partial S}\,\dd S+\nu\,\dd\hat\sigma,
 \label{eq:optionfirst}
\end{equation}
with gamma, vanna, volga and jump risks retained when material.
Empirical option-impact estimates provide useful magnitudes but do not by
themselves identify the risk and time exponents \cite{Toth2016}.

Let $q_j$ denote the signed vega exposure accepted by an LP in option
bucket $j$. Choose fixed factor loadings $B_{mj}$, normalised so that the
factor exposures $z_m=\sum_j B_{mj}q_j$ are conjugate to factor quote
changes; the factors may include volatility level, skew, curvature, and
term-structure slope. Suppose each \emph{hedged residual mode} has a risk-scaling
law of the form \eqref{eq:riskscaling}, with its own
$\beta_m,H_m,\gamma_m$. 
Applying the single-inventory pricing rule separately to each residual mode gives the quote adjustment for factor $m$,
\begin{equation}
 h_m(z_m)=c_m\sgn(z_m)|z_m|^{\delta_m},\qquad
  \delta_m=1-\beta_m+\gamma_m(2H_m-1).
 \label{eq:factormodes}
\end{equation}
Here \(c_m>0\) incorporates the exposure-unit normalisation. Summing the factor adjustments with their loadings gives the implied-volatility quote adjustment for bucket $j$,
\begin{equation}
 \Delta\hat\sigma_j=\sum_m B_{mj}h_m(z_m).
 \label{eq:surface}
\end{equation}
A position at one strike can therefore affect the quote at another through their shared factor exposures.


The Hurst exponent
of quoted log-volatility \cite{GJR2018} need not equal that of the LP's
hedged residual P\&L. And near expiry, vega may become small while
delta, settlement, and gap exposures dominate, so a pure vega coordinate
must be supplemented by spot and settlement factors. 

\section{Application II: The Overnight Anomaly and Institutional Clocks}
\sectionstanza{
``But not on us!'' the Oysters cried,\\
\stanzaindent Turning a little blue.\\
``After such kindness, that would be\\
\stanzaindent A dismal thing to do!''\\
``The night is fine,'' the Walrus said.\\
\stanzaindent ``Do you admire the view?'' 
}
\label{sec:knut}
Knuteson's observation of the historical overnight--intraday return split
\cite{Knuteson2020,Knuteson2022} has been widely discussed.
Buying near the open and
differences in trading clientele across the day have also been documented
\cite{Berkman2012,LouPolkSkouras2019}. Consider a broker\footnote{A VWAP broker executes
for a fee and is assessed against completion and the benchmark; an LP
trades as principal and may hedge across assets and seek statistical
inventory gains. These roles have long coexisted within firms.} executing a
client order against the day's volume-weighted average price (VWAP), with
completion required by the close. 
Automatic retirement contributions and default investments
can generate such orders \cite{IRS1998,PPA2006,DOLQDIA2007,Choi2024}. 
The broker chooses when to trade, before the day's volume
profile is known \cite{BussetiBoyd2015,McCullochKazakov2007}.

If more volume than the broker expects arrives subsequently, then waiting and 
trading at a lower participation rate is preferable.
 If less volume arrives,
the broker must execute the outstanding shares in a thinner market.
These forecast errors have different costs. For a given number of shares
left to trade, a shortfall in market volume raises the cost of completion
by more than an equally large surplus reduces it.
To see this\footnote{
Kelly betting exhibits similar behaviour, since uncertainty about the `edge' favours  smaller positions (Appendix~\ref{app:growth}).
}, divide the session into an opening window with known market
volume $V_O>0$ and a later window with uncertain volume $V_L>0$. Take
these volumes as given by the rest of the market. The client order has unsigned
size $Q_i>0$, of which the broker executes $x$ shares in the opening window.
Suppose the average impact cost per share\footnote{ 
Under constant-share aggregation, the cost exponent can be identified with the LP's impact exponent, $\alpha=\delta$, when the same power-law concession governs execution costs.
%
} when trading $u$ shares against
volume $V$ is $c(u/V)^\alpha$, with $c>0$ and $\alpha>0$ common to both
windows. 

For the broker’s timing decision, take the reference price to be the same in both windows and the market VWAP to be unaffected by this client’s execution schedule.
The broker then minimises expected execution cost,
\[
 C(x)=c\,\frac{x^{1+\alpha}}{V_O^\alpha}
      +c\,(Q_i-x)^{1+\alpha}\E[V_L^{-\alpha}],
 \qquad 0\le x\le Q_i,
\]
where expectations are conditional on information available when $x$
is chosen and the moments shown are finite. Equating marginal costs
across the two windows gives the optimal opening fraction
\begin{equation}
 \frac{x^*}{Q_i}
 =\frac{V_O}{V_O+\bigl(\E[V_L^{-\alpha}]\bigr)^{-1/\alpha}}
 >\frac{V_O}{V_O+\E[V_L]}, 
 \label{eq:knutmv}
\end{equation}
since for non-constant $V_L$, Jensen's inequality gives
$\E[V_L^{-\alpha}]>(\E[V_L])^{-\alpha}$.
Even a risk-neutral broker therefore executes a larger fraction in the
opening window than it would if late volume were certain to equal its
forecast mean\footnote{
A broker that penalises execution-cost variance chooses a larger opening allocation than the expected-cost optimum in \eqref{eq:knutmv}, since this reduces exposure to uncertain late volume. With the benchmark fixed, the
variance of realised slippage per share is
$c^2(Q_i-x)^{2+2\alpha}\Var(V_L^{-\alpha})/Q_i^2$, assuming
$\E[V_L^{-2\alpha}]<\infty$, and moving a marginal share forward reduces this variance
while leaving expected cost unchanged to first order.}.

The timing incentive applies to both buys and sells. We conjecture that clients seeking execution at the open are, on balance, buyers: a fund investing pension contributions may face a tighter deadline than shareholders willing to sell. Let $Q_O$ denote net customer buying at the opening price, measured as shares bought minus shares sold. If a marginal LP absorbs this imbalance, its quote satisfies
\begin{equation}
 p_O=K_O+A\,\sgn(Q_O)\left|Q_O/q_*\right|^\delta .
 \label{eq:knutopen}
\end{equation}
Here $K_O$ is the LP’s reference value\footnote{Liquidity providers need not agree on a reference value.} before allowing for inventory risk. When $Q_O>0$, the LP sells at a premium and bears inventory risk while waiting for offsetting sellers. The price at which it can cover remains uncertain. Define the log premium at time $t$ as $\log(p_t/K_t)$. If the expected log premium is higher at the recorded open than at both the preceding and subsequent closes, it contributes positively to close-to-open returns and negatively to open-to-close returns.
Together, these mechanisms shed light on the overnight-intraday return anomaly.

\section{Conclusion}
\sectionstanza{
``O Oysters,'' said the Carpenter,\\
\stanzaindent ``You've had a pleasant run!\\
Shall we be trotting home again?''\\
\stanzaindent But answer came there none---\\
And this was scarcely odd, because\\
\stanzaindent They'd eaten every one.
}
\label{sec:scope}
The paper extends the earlier risk–return pricing argument for liquidity providers~\cite{Meister2026}. While actual wealth is locally fixed, both the risk of the rebalanced residual portfolio and its holding horizon may vary with position size. These two scalings yield a general impact exponent and a condition for increasing, concave supply. Incorporating the size-dependent horizon directly into the optimisation changes the coefficient, not the exponent.

The relation derived here gives the LP’s quote exponent for its current risk and holding-time scalings. These scalings may vary together\footnote{Treating $\beta$, $H$ and $\gamma$ as inputs does not require them to be independent.} and may differ across markets, so commodity and equity books can give rise to different quote exponents. Over time, the LP’s quotes may themselves shape the order flow it attracts and the risk of the resulting book. Whether this feedback selects a unique stable outcome, permits several, or sustains fluctuations in the exponents remains open.
Consider the portfolio exponent $\beta$. For each quote, the LP takes $\beta$ as given. It describes how the incremental variance of the existing book scales with the size of the proposed trade, allowing for hedging and netting under the LP’s capital constraints. Neither the stock’s OU dynamics nor the quadratic criterion determines it.
Identifying a supply schedule with observed impact requires the
LP's quote to be marginal; market volume and inter-LP feedback remain
outside the model. 

 Liquidity supply need not be passive. LPs are known to pay brokers
for retail flow~\cite{EasleyKieferOHara1996}. Low average information content of this flow
does not imply independent arrivals: customers without private signals may
still exhibit herding. 
The LP need not possess private information about value to
price the resulting inventory, but it must assess the expected risk and
return. 

A market does not require agreement on the value of an asset, only agreement between counterparties on the price at which they are willing to exchange it. Investors may differ in their valuations, intertemporal discounting, tax treatment and institutional constraints. Price discovery need not be the discovery of a common value.

Market clearing, however, does not necessarily imply market efficiency. Individually rational behaviour under institutional constraints can produce persistent opportunities for excess returns~\cite{Meister2023}. These need not arise from superior information about value but may instead result from the market mechanism itself. Market inefficiency may emerge from the rules of the game rather than from dubious conduct of the players.

 The Oysters expect a winsome exchange; the Walrus sees a business.




\appendix
\section{Limits of the Quadratic Growth Calculation}
\label{app:growth}
\sectionstanza{
``If seven maids with seven mops\\
\stanzaindent Swept it for half a year,\\
Do you suppose,'' the Walrus said,\\
\stanzaindent ``That they could get it clear?''\\
``I doubt it,'' said the Carpenter,\\
\stanzaindent And shed a bitter tear.}

The expected-log-growth criterion of~\cite{Meister2026} is the starting
point. Take $t=0$ as the decision time, let $W=W_0>0$ be the LP's current
wealth, and write $\E_0$ for expectation given the information available
at that time.
Let $Y=W_T-W$ be the gain over the holding horizon, with $W+Y>0$.
The certainty equivalent of $Y$ is the sure gain $c$ that gives the LP
the same expected utility, so $\log(W+c)=\E_0[\log(W+Y)]$.
Solving for $c$ gives
\[
 c=W\left\{\exp\!\left(\E_0[\log(1+Y/W)]\right)-1\right\}.
\]
Provided the logarithmic series converges and can be averaged term by term,
\begin{equation}
 \E_0[\log(W+Y)-\log W]
  =\E_0[Y]/W-\E_0[Y^2]/(2W^2)
   +\E_0[Y^3]/(3W^3)-\cdots.
 \label{eq:kellytaylor}
\end{equation}
 Expanding the exponential in the expression for $c$ to second order
in $Y/W$ gives
\[
 c \approx \E_0[Y]
 -\frac{\E_0[Y^2]-(\E_0[Y])^2}{2W}
 =\E_0[Y]-\frac{\Var_0(Y)}{2W}.
\]
Subtracting the certainty equivalent of the existing book from that
of the book after the trade gives \eqref{eq:CE} to second order
under the stated gain forecast.

 With $q$ shares, its self-financing wealth increment is $\mathrm dW_t=q(\mu\,\mathrm dt+\sigma\,\mathrm dB_t)$. It\^o's formula gives the instantaneous expected log-growth rate at $t=0$,
\begin{equation}
 g=\frac{q\mu}{W_0}-\frac{q^2\sigma^2}{2W_0^2}.
 \label{eq:ito}
\end{equation}
Higher-order terms do not contribute to this instantaneous continuous-diffusion rate, though they may matter over discrete periods or in the presence of jumps.

 A catastrophe bond with
gain $r$ upon survival and loss of principal upon default has one-period
expected log growth
\begin{equation}
 (1-p)\log(1+fr)+p\log(1-f),
 \label{eq:cat}
\end{equation}
where \(p\) is the default probability and $f$ is the invested fraction. This expression cannot generally be replaced by a quadratic in $f$, as discussed in earlier papers~\cite{ LvMeister2010, Meister2026}. 
Using this second-order approximation over the range of inventories considered is a modelling assumption.

Eq.~\eqref{eq:CE} is a definition; two utility-based certainty equivalents are related to it as follows. Let $u(w)$ be a utility function of terminal wealth $w$, with absolute risk aversion $-u''(w)/u'(w)$. For log utility this is $1/w$, which equals $1/W$ at the LP's current wealth $W$; for exponential utility with risk tolerance $W$ it equals $1/W$ at every $w$. To second order, the certainty equivalent of a small gain $Y$ depends on the utility only through this local risk aversion, as $\E[Y]-\Var(Y)/(2W)$, so the two utilities agree through second order in $Y/W$ and differ beyond it. If terminal portfolio P\&L is conditionally Gaussian before and after the trade, the exponential-utility certainty equivalent is exactly $\E[Y]-\Var(Y)/(2W)$, and its change over the trade is \eqref{eq:CE}. This reading needs no It\^o formula, which does not apply to fractional Brownian motion with $H\ne\tfrac12$, but it changes the utility function and does not justify the log-utility criterion beyond second order.

\paragraph{Departures from the quadratic criterion.}
Let $f$ be the invested wealth fraction, and $f_{\rm MV}$ the maximiser of the
quadratic criterion \eqref{eq:CE}. That criterion drops the terms of
\eqref{eq:kellytaylor} beyond second order. This loses nothing in the Gaussian,
exponential-utility case above, nor in the instantaneous growth rate
\eqref{eq:ito} of a continuous diffusion. Expected log growth  over a finite horizon has a maximiser $f^*$ and, because of the logarithm, falls faster above it than below it. Departures from $f_{\rm MV}$ emerge when adjustment is discrete or the optimum is uncertain:
(i)~\emph{Discrete adjustment.} If the position can be changed only at intervals of length $\tau_{\rm adj}$,
the gain between adjustments is not infinitesimal, so the terms of \eqref{eq:kellytaylor} beyond second order matter and, for Gaussian P\&L under log utility, shift the optimum downward. The shift vanishes as $\tau_{\rm adj}\to0$.
(ii)~\emph{Uncertainty about the optimum.}
Because of the asymmetry, overshooting
$f^*$ costs more than undershooting, 
so when $f^*$ is uncertain and only a point estimate (the mean), here  $f_{\rm MV}$, is available, it is better to err on the side of caution.
In both cases the optimal exposure lies below $f_{\rm MV}$.

\end{document}